# The role of surface chemical reactivity in the stability of electronic nanodevices based on two-dimensional materials "beyond graphene" and topological insulators

A. Politano[a,*], M.S. Vitiello[b], L. Viti[b], D.W. Boukhvalov[c,d], G. Chiarello[a]

a Department of Physics, University of Calabria, via ponte Bucci, 31/C, 87036 Rende (CS), Italy
b NEST, Istituto Nanoscienze–CNR and Scuola Normale Superiore, Piazza San Silvestro 12, Pisa I-56127, Italy
c Department of Chemistry, Hanyang University, 17 Haengdang-dong, Seongdong-gu, Seoul 133-791, Republic of Korea
d Theoretical Physics and Applied Mathematics Department, Ural Federal University, Mira Street 19, 620002 Ekaterinburg, Russia

*Here, we examine the influence of surface chemical reactivity toward ambient gases on the performance of nanodevices based on two-dimensional materials "beyond graphene" and novel topological phases of matter. While surface oxidation in ambient conditions was observed for silicene and phosphorene with subsequent reduction of the mobility of charge carriers, nanodevices with active channels of indium selenide, bismuth chalcogenides and transition-metal dichalcogenides are stable in air. However, air-exposed indium selenide suffers of p-type doping due to water decomposition on Se vacancies, whereas the low mobility of charge carriers in transition-metal dichalcogenides increases the response time of nanodevices. Conversely, bismuth chalcogenides require a control of crystalline quality, which could represent a serious hurdle for up scaling.*

Corresponding author: Dr. Antonio Politano

e-mail: antonio.politano@fis.unical.it

Tel. +39-0984-496107 Fax +39-0984-494401

In the last decade, graphene had a giant impact on scientific research [1]. However, for applications requiring a switching of the conductivity, the use of graphene is not appropriate since it has not a band gap [2]. Field-effect transistors necessitate a large value of the ON/OFF ratio, which is defined as the ratio of currents in the on- and off-state of the nanodevice [3, 4]. For this reason, recently many other classes of two-dimensional (2D) materials "beyond graphene" have emerged [5, 6].

Nature provides a huge variety of layered materials "beyond graphene" (semimetals, semiconductors, insulators) [7] with electronic band gaps which span from the infrared to the ultraviolet. The synthesis of novel 2D materials, such as transition-metal dichalcogenides (TMDC), 2D carbides, IV-VI compounds or atomically thin elemental materials (silicene, germanene, phosphorene) promises a revolutionary step-change, since they exhibit exotic physicochemical properties, which have never been accessed so far with three-dimensional materials. These innovative 2D materials enable the combination of flexibility and transparency with an existing electronic band gap.

In 2010 it has been claimed that silicene, a 2D layer of silicon obtained by depositing Si on an Ag(111) single crystal [8], could host massless Dirac fermions with ultrahigh mobility [9]. The strong chemical bonds formed by silicene with different molecules ($NH_3$, $NO$, $NO_2$) makes it a promising material for the design of molecular sensors [10]. However, silicene is not stable in ambient conditions: silicene oxidation is favored even at relatively low $O_2$ doses [11]. In principle, the adsorption of oxygen atoms could be used for band gap tuning [12], but a moderately oxidized silicene monolayer on Ag(111) is semimetallic [13]. The emergence of a Si-O-Si configuration in the early phases of silicene oxidation implies a structural rearrangement, which further accelerates the oxidation process at high oxygen doses [11]. At the completion of the oxidation process, the silicene structure is replaced by a $sp^3$-like tetrahedral configuration Silicene oxidation can be prevented by using capping layers [14]. Nevertheless, silicene suffers of other impediments. As a matter of fact, researchers have also pointed out that spectral features previously attributed to Dirac-cone electrons in silicene are actually silver-related [15]. Moreover, the lack of a parental bulk crystal from which exfoliating ultrathin flakes of silicene represents a severe obstacle for the nanofabrication process. Thus, one can conclude that a silicene-based technology has few possibilities to be effectively realized. Similar considerations also apply for germanene [16].

The achievement of ON/OFF ratios as high as $10^8$ in 2D TMDC ($MoS_2$, $WS_2$, $MoSe_2$, $WSe_2$) [17], as well as their good ambient stability could in principle provide an interesting pathway. However, mobility of charge carriers in TMDC is about three orders of magnitude less than in graphene with a subsequent increase of the response time of any TMDC-based nanodevice [18].

The high mobility of charge carriers in phosphorene makes it a solid candidate for nanoelectronics [19]. Phosphorene is a single layer of black phosphorus, the most stable allotrope of phosphorus at ambient conditions. Phosphorene is a nonplanar and anisotropic material, with additional degree of freedom for nanodevices [20]. It has been predicted to be suitable for the design of selective and sensitive gas sensors for detecting most common ambient gases [21]. Such predictions are also supported by experimental evidence of stable CO adsorption at room temperature on phosphorene [22]. However, phosphorene should be protected by a capping layer [23] in order to avoid the formation of surface $P_2O_5$ species [24], which inevitably cause a notable decrease of the mobility of charge carriers [25] and surface degradation in environmental conditions [26]. In fact, the presence of lone pairs in P atoms of phosphorene results in higher reactivity [27]. Figure 1 reports the evolution of the morphology of phosphorene flakes in ambient conditions. Both the atomic force microscopy (AFM) images in panel (a) and microwave impedance microscopy maps in panel (b) indicate an apparent reduction of the thickess due to surface degradation. Once encapsulated, the mobility of phosphorene-based nanodevices is constant in time [25] (panel c of Figure 1). Even if the stabilization of phosphorene layers in ambient conditions has been recently claimed by several researchers [28-30], the reliability of such approaches for up scaling is complicated and, consequently, the practical use of phosphorene in technology is hitherto elusive.

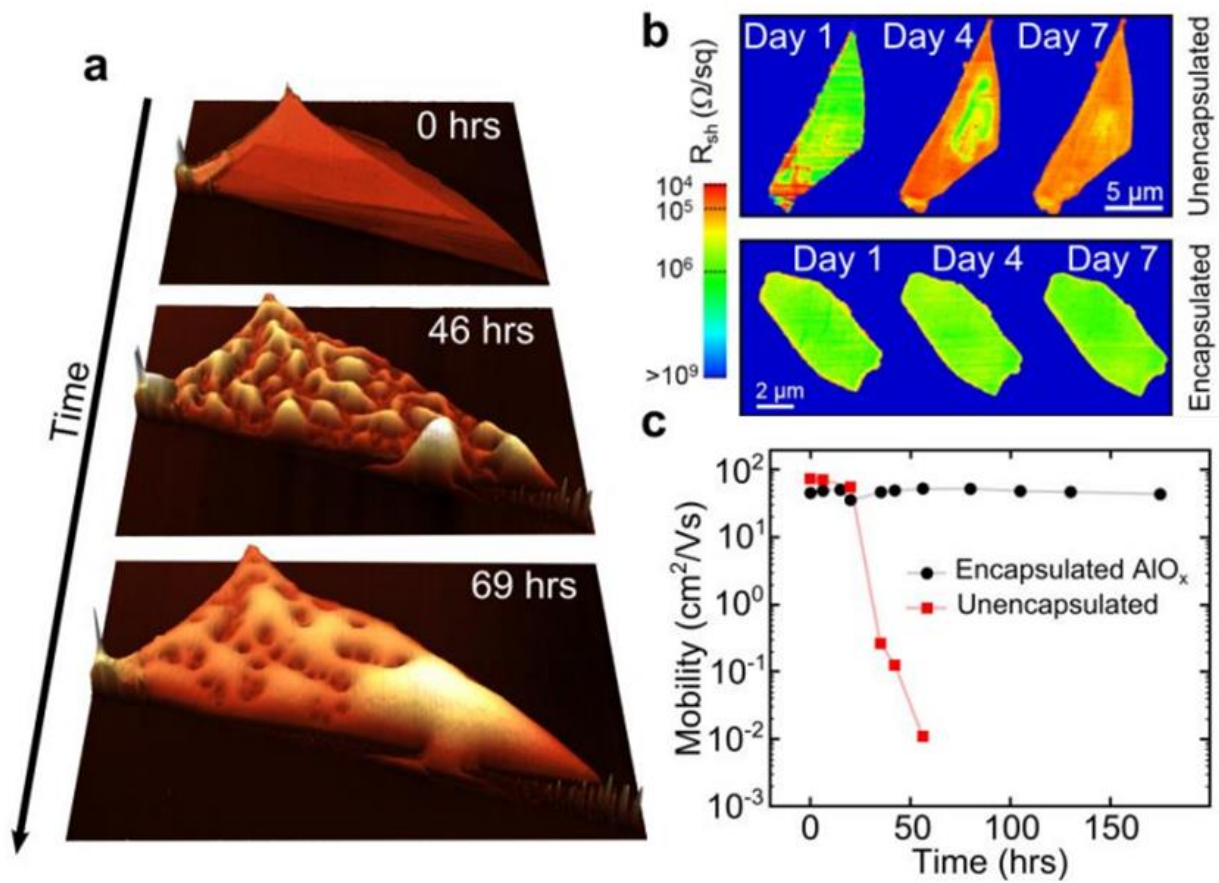

*Figure 1. (a) AFM images of a black-phosphorus flake left in air for 0, 46, and 69 hours after exfoliation (data taken from Ref. [26]). (b) Microwave impedance microscopy map of the local resistance of a couple of black-phosphorus flakes: one unencapsulated and another encapsulated with a thick film of aluminum oxide. It can be noticed that the oxidation starts in correspondence of the edges of the uncapped black-phosphorus flake. Data have been taken from Ref. [31] (c) Time evolution of the mobility of charge carriers in phosphorene-based field-effect transistors. Without the use of a capping layer to preserve the performance of the nanodevice, a rapid degradation in air is observed (data taken from Ref. [25]).*

*Panels a-c have been adapted with permission from Ref. [32]. Copyright (2015) American Chemical Society.*

Another suitable candidate for nanoelectronics is represented by InSe, which is a layered semiconductor made of stacked layers of Se-In-In-Se atoms with van der Waals bonds between quadruple layers [33, 34]. Recently, the outstanding performance of InSe-based optoelectronic devices have been reported by many groups [35, 36]. Field-effect transistors with an active channel of InSe are characterized by an electron mobility near $10^3$ cm$^2$/(V s) [35] and, additionally, excellent flexibility [37, 38] and ambient stability [39], in spite of the presence of a *p*-type doping arising from water decomposition at Se vacancies [39]. By contrast, no reactivity has been found toward oxygen [39]. The ambient stability of the surface of InSe is important for its photovoltaic applications [37].

The recent rise of topological phases of matter affords other candidates for nanoelectronics and optoelectronics. In particular, topological insulators (TIs) combine the presence of bulk band gap with spin-polarized surface states forming a Dirac cone as for graphene [40]. Their peculiar band structure enable (*i*) the design of spin-valve devices [41] and (*ii*) the reduction of low-frequency noise in TI-based electronic devices [42, 43]. Among TIs, bismuth chalcogenides have the highest application capabilities [44-48] and, consequently, this class of TIs has attracted great attention [49].

Topological surface states are found to exist in crystals [50] and films [51] possessing high structural quality (both stoichiometric/ composition and crystallographic). The presence of vacancies shifts the position of the Fermi level, altering the properties of the TI system [52]. Moreover, the presence of Se vacancies favours the formation of the Bi-O bond [53] and the rapid surface oxidation and degradation of Bi$_2$Se$_3$ [54]. The use of the Bridgman-Stockbarger method somewhat suppresses the quantity of vacancies in TI samples [55]. This implies chemical inertness toward ambient gases [53, 55, 56] (see Figure 2a and its caption) and the possibility to tune the position of the Fermi level [57, 58]. High-quality single crystals of TIs, grown by the Bridgman-Stockbarger method, have been recently used as active channels of field-effect transistors for Terahertz photodetection (Figure 2b), which resulted into an exceptional air stability and outstanding detection performances [44] (inset of Figure 2b). However, the extreme control of the crystalline quality required for TIs is unsustainable for up scaling.

Following TIs, additional novel topological phases of matter, such as Dirac [59] and Weyl semimetals [60], have recently appeared. Both Dirac ($Cd_3As_2$, $Na_3Bi$) and Weyl (NbP, TaP, NbAs, TaAs) semimetal are characterized by a giant value of the magnetoresistance [61-64] and by ultrahigh mobility of charge carriers [62, 63, 65], even larger than that of graphene. Their hitherto unexplored surface chemical reactivity may unveil novel capabilities and/or pitfalls for the exploitation of the novel topological phases of matter in nanoelectronics.

Similarly, investigations of surface chemical reactions at van der Waals heterostructures formed by basic building blocks of 2D materials [66] are still missing.

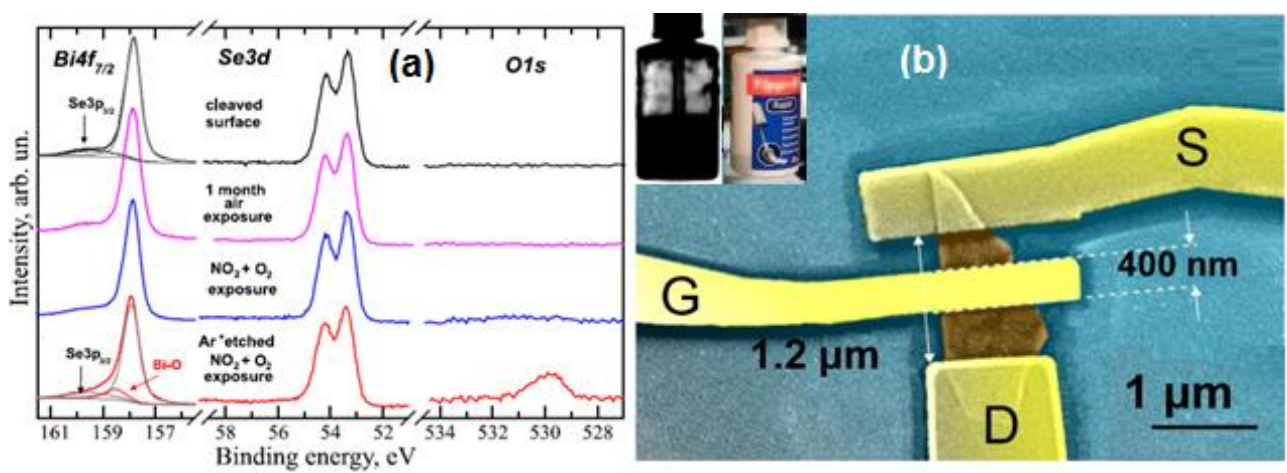

*Figure 2. (a) XPS spectra in the region of the Bi $4f_{7/2}$, Se 3d, and O 1s core levels acquired for $Bi_2Se_3$ (black curve) as-cleaved, (pink curve) after one month in air, (blue curve) after exposure of freshly cleaved surface to a dose of 5 kL of $NO_2+O_2$ and (red curve) after bombardment with Ar ions followed by a dose of 3 kL of $NO_2+O_2$. The O 1s core level has been recorded only in defected samples. Adapted with permission from Ref. [53]. Copyright (2012) American Institute of Physics. (b) Field-effect transistor fabricated with unencapsulated $Bi_2Te_{2.2}Se_{0.8}$ (S, G, D represent the source, gate and drain electrodes). An imaging experiment with Terahertz radiation on a jar containing glue carried is reported in the inset. Adapted with permission from Ref. [44]. Copyright (2016) American Chemical Society.*